# The insulating phases of vanadium dioxide are Mott-Hubbard insulators

T. J. Huffman[1], C. Hendriks[1], E. J. Walter[1], Joonseok Yoon[2], Honglyoul Ju[2], R. Smith[3],

G. L. Carr[3], H. Krakauer[1], M. M. Qazilbash[1,*]

1. *Department of Physics, College of William and Mary, Williamsburg, VA 23187-8795, USA*
2. *Department of Physics, Yonsei University, Seoul 120-749, Republic of Korea*
3. *Photon Sciences, Brookhaven National Laboratory, Upton, NY 11973, USA*

**We present the first comprehensive broadband optical spectroscopy data on two insulating phases of vanadium dioxide ($VO_2$): monoclinic $M_2$ and triclinic. The main result of our work is that the energy gap and the electronic structure are essentially unaltered by the first-order structural phase transition between the $M_2$ and triclinic phases. Moreover, the optical interband features in the $M_2$ and triclinic phases are remarkably similar to those observed in the well-studied monoclinic $M_1$ insulating phase of $VO_2$. As the energy gap is insensitive to the different lattice structures of the three insulating phases, we rule out Peierls effects as the dominant contributor to the opening of the gap. Rather, the energy gap arises from intra-atomic Coulomb correlations.**

There have been many experimental and theoretical studies of the thermally driven metal-insulator transition (MIT) between the insulating monoclinic ($M_1$) and the metallic rutile (R) phases of vanadium dioxide ($VO_2$). Some fraction of these studies attribute the insulating $M_1$ state to the vanadium-vanadium Peierls type pairing (see Fig. 1) that leads to unit cell doubling. Others argue that the insulating behavior in the $M_1$ phase is primarily a result of Mott-Hubbard correlations. These studies, reviewed in Refs. [1–3], are too numerous to be referenced here. A significant proportion of the literature on the nature of insulating $VO_2$, particularly in recent years [4–11], has struggled to decouple the contributions of the Mott-Hubbard and Peierls mechanisms because of an emphasis on the $M_1$ phase. Interestingly, it has long been recognized that measuring the electronic properties of two additional insulating $VO_2$ phases, the monoclinic $M_2$ and triclinic T, could potentially settle the debate about the origin of the energy gap, but the measurements have been difficult to achieve.

One of our purposes in this Letter is to refocus attention to the importance of measuring the electronic properties of the monoclinic $M_2$ and triclinic (T) phases to decouple the effects of the Peierls and Mott-Hubbard mechanisms. This can be seen from the argument put forward by Pouget *et al.* [12], which can be summarized as follows: One starts from a model of an isolated vanadium dimer in $VO_2$, with one electron per site, analogous to the familiar case of the hydrogen molecule. Both the Peierls and Mott-Hubbard pictures correspond to limiting cases of the Hubbard model for a chain of such dimers, depending on whether the intra-dimer hopping parameter ($t$) or the intra-atomic Coulomb repulsion ($U$), respectively, is the dominant energy scale in the system. Interestingly, in both cases, the qualitative description of the electronic structure is the same: an insulator with a bonded spin singlet on the dimer, where the band gap results from splitting of the bonding and anti-bonding $a_{1g}$ bands (the lower and upper Hubbard bands in the Mott picture). As pointed out by the authors of Ref. [12], the only clear distinction between the two cases is how the energy gap responds to changes in the hopping parameter resulting from changes in lattice structure. For the chain of dimers, the bands broaden relative to the isolated dimer, decreasing the gap based on the inter-dimer hopping ($t'$). In the Peierls limit ($U \ll t, t'$), insulating behavior vanishes as $t'$ approaches $t$, the case of undimerized chains. In contrast, the gap is primarily set by $U$ in the Mott-Hubbard limit ($U \gg t, t'$), and thus insensitive to changes in the degree of dimerization. In the $M_1$ phase, where all of the chains are dimerized and equivalent, it is impossible to decouple the effect of dimerization from intra-atomic Coulomb correlations. This is not the case for the $M_2$ and T phases.

In this Letter, we present broadband optical spectroscopy data on the $M_2$ and T phases of $VO_2$. We have performed infrared micro-spectroscopy and spectroscopic micro-ellipsometry on internally strained $VO_2$ crystals that undergo a first order phase transition with increasing temperature from the T phase to the $M_2$ phase. The energy gap and electronic structure are essentially unchanged across this structural phase transition. Moreover, the optical energy gap of 0.6 ($\pm$ 0.1) eV in the $M_2$ and T phases is nearly the same as that measured by numerous previous measurements on the $M_1$ phase [10,13–17]; the gap is insensitive to the different vanadium pairing arrangements in the $M_1$, $M_2$ and T phases. It follows that the gap has a common physical origin in the intra-atomic Coulomb correlations in the insulating phases of $VO_2$. This conclusion is supported by calculations also presented in this work.

In the $M_1$ phase, all of the vanadium ions dimerize and tilt in equivalent chains along the rutile $c_R$ axis (see Fig. 1). In contrast, the $M_2$ phase contains two types of vanadium chains: one type consists of vanadium ions that pair but do not tilt, while the other consists of vanadium ions that tilt but do not pair. The vanadium ions in the latter chain are equidistant, each carrying a localized electron with a spin-1/2 magnetic moment and antiferromagnetic exchange coupling between nearest neighbors [12]. The T phase has two types of inequivalent vanadium chains (or sub-lattices) in which the vanadium ions are paired and tilted to different degrees (see Fig. 1.) [12,18]. The T phase can be thought of as an intermediate phase between the $M_2$ and $M_1$ phases, where the chains become equivalent in $M_1$. While the $M_1$ insulating phase is generally found in bulk $VO_2$, the $M_2$ and T phases can be accessed via chemical doping or strain [12,19–25].

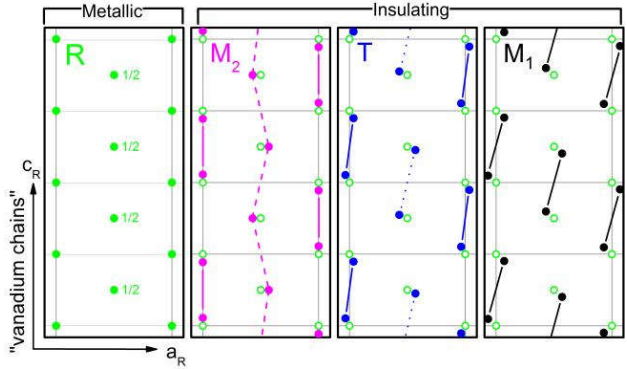

FIG. 1. A plan view of vanadium ion positions for the metallic rutile and insulating $M_2$, T, and $M_1$ phases of $VO_2$. In all phases, the vanadium ions at the center of each rutile unit cell (shown by the gridlines) are offset from the others by ½ unit cell (denoted by "1/2" in the rutile panel). The vanadium ions in the insulating phases undergo small displacements from the rutile positions (open green circles in the panels of the insulating phases). The rutile lattice vectors $c_R$ and $a_R$ are shown in the lower left corner of the diagram. Vanadium chains in the insulating phases are oriented along the $c_R$ direction.

Nuclear Magnetic Resonance (NMR) and Electron Paramagnetic Resonance (EPR) have determined the presence of localized d-electrons with about one Bohr magneton magnetic moment on the unpaired vanadium chains of the $M_2$ phase [12,21,22]. While this localization is a clear hallmark of a Mott-Hubbard insulator, the situation in the dimerized chains is less clear. The NMR and EPR measurements reveal that the electrons on the dimerized chains are covalently bonded. Therefore, as alluded to above, it is ambiguous whether the dimerized chains should be thought of as Peierls insulators, or Mott-Hubbard insulators with the valence electrons forming covalently bonded singlets which are localized on the dimers. It has been argued that the $M_1$ and T insulating phases of $VO_2$, which differ only slightly in free energy from the $M_2$ phase, cannot have a grossly different energy gap and should thus also be classified as Mott-Hubbard insulators [12,26]. Although strong, this argument needs experimental verification – presented here– via direct measurement of the $M_2$ and T phase energy gaps, which can then be compared to each other and to the literature values of the $M_1$ phase energy gap.

Pure $VO_2$ crystals were grown with a self-flux method and thoroughly characterized with transport and X-ray diffraction measurements [23,27]. X-ray diffraction together with resistance measurements have determined that upon heating, the crystal first goes through an insulator-to-insulator transition and a structural transition between the T phase and the $M_2$ phase [23,27]. The temperature dependent resistance data displayed in the inset of Fig. 2 clearly shows two discontinuities along with hysteresis, indicative of first order phase transitions. The resistance increases by about a factor of two upon the structural transition from the T phase to the $M_2$ phase, consistent with previous measurements in the literature [25]. Upon further heating, the crystal undergoes an insulator to metal transition and a structural change from the $M_2$ phase to the metallic R phase. Unpolarized Raman micro-spectroscopy on the T, $M_2$, and R phases, presented in Fig. 2, verifies the structural assignment from x-ray diffraction when compared to unpolarized Raman spectra in the literature [20,28]. Our crystals are in the shape of rods with approximately square cross-sections between 50 μm and 100 μm wide, and with lengths between 1 mm and 3 mm. The rutile $c_R$ axis, which points along the vanadium chains in the insulating phases, is oriented along the long axis of the crystals. Through an optical microscope, we observe that the crystal increases in length by ~ 0.6% upon transitioning from the T phase to the $M_2$ phase, and decreases in length by ~ 1.7 % across the MIT from $M_2$ to R. These changes in length are consistent with the changes in the lattice parameters along the vanadium chains measured with x-ray diffraction in previous works [29,30]. The surface of the crystal is identified by X-ray diffraction as the (110) plane in the rutile basis which transforms to two coexisting, twinned surfaces (201) and ($\bar{2}$01) in the monoclinic $M_2$ phase [31]. Further twinning occurs as the 2-fold rotational symmetry of the $M_2$ phase is lost upon transitioning to the T phase. The result is that for each $M_2$ twin, there are two possible T phase twins, which differ from each other by a 180° rotation along the $c_R$ ($b_{M2}$) direction [32].

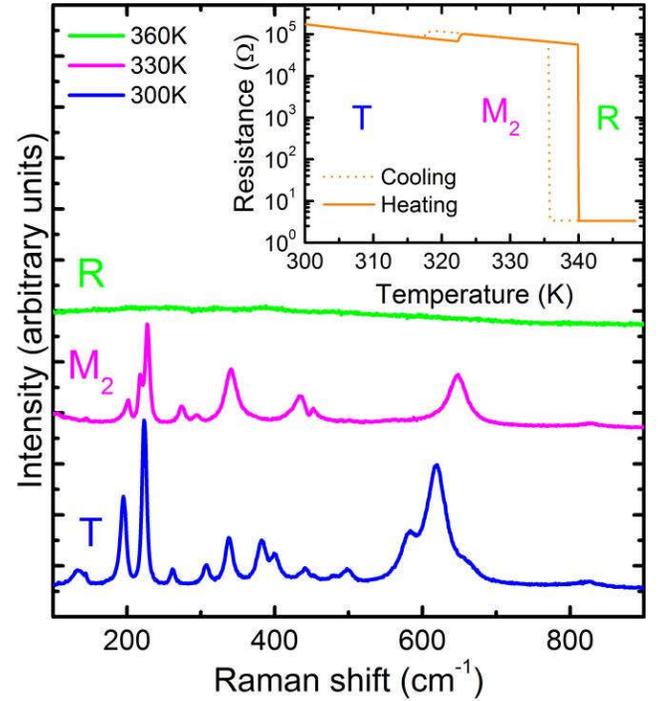

FIG. 2. Raman spectra of a typical $VO_2$ crystal studied in our work. Spectra are shown for the T, $M_2$ and R phases. The *dc* resistance of the crystal is plotted as a function of temperature in the inset.

The small size of the $VO_2$ crystals calls for specialized infrared and optical micro-spectroscopy techniques to obtain reliable data with good signal-to-noise ratio. Infrared reflectance micro-spectroscopy between 150 and 6000 $cm^{-1}$ was performed at beamline U12IR at the National Synchrotron Light Source, Brookhaven National Laboratory [33]. Infrared polarizers were employed to obtain reflectance spectra parallel and perpendicular to the long axis of the crystals, i.e. the rutile $c_R$ direction. Absolute values of the infrared reflectance spectra in the T and $M_2$ insulating phases were obtained by normalizing them to the nearly featureless spectra of the rutile metal.

Generalized spectroscopic micro-ellipsometry between 0.6 and 5.5 eV (~4800 and ~44000 $cm^{-1}$) was performed at William and Mary using an in-house focusing set-up coupled to a Woollam Variable Angle

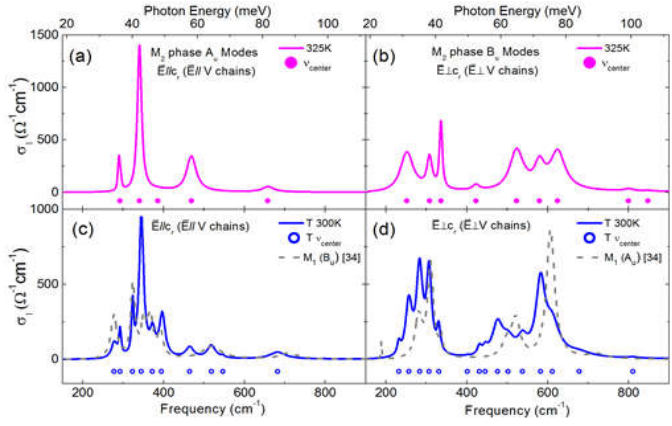

FIG. 3. Polarization dependent optical conductivity ($\sigma_1$) showing the infrared active phonon spectra of the $M_2$ phase (panels (a) and (b)), and the T phase (panels (c) and (d)). The center frequencies of the phonon features are denoted by circles labeled $\nu_{center}$. The previously reported $M_1$ phase infrared active phonon spectra [34] are compared to the triclinic phase spectra in (c) and (d).

Spectroscopic Ellipsometer (W-VASE). Spectroscopic ellipsometry has the notable advantage over reflectance spectroscopy alone in that it preserves information related to the phase shift upon reflection, enabling the accurate determination of both the real and imaginary parts of the optical constants of the material. Data for three angles of incidence was obtained on the crystals oriented with their long axis parallel and perpendicular to the plane of incidence. The same crystals and heating arrangement were used for both the reflectance and ellipsometry experiments. All data sets were analyzed together in the W-VASE software with Kramers-Kronig consistent oscillators to obtain the broadband, frequency dependent complex conductivity parallel and perpendicular to the vanadium chains in the T and $M_2$ phases.

We present the first report on polarization dependent optical conductivity data on the infrared-active phonons of $M_2$ and T phases in Fig. 3. For the $M_2$ phase (space group $C2/m$), group theory predicts 6 $A_u$ phonon modes for light polarized parallel to the $b_{M_2}$ axis oriented along the vanadium chains, and 9 $B_u$ phonon modes for light polarized perpendicular to the $b_{M_2}$ axis. We observe 5 $A_u$ and all 9 $B_u$ phonon modes in the experimental spectra. It is possible that the sixth $A_u$ phonon mode has a weak dipole moment and therefore is not seen in experiment. The discontinuous structural phase transition to the T phase is captured by the significant increase in the number of phonon features in the T spectra in both polarizations. This is explained by the lower symmetry of the triclinic structure (space group $C\bar{1}$). In Fig. 3, we include the $M_1$ spectra from our previous work for comparison [34]. The $M_1$ phonon spectra resemble the T phonon spectra and lead to the conclusion that the T phase is merely a slight structural distortion of the $M_1$ phase. Indeed this is consistent with past observation of the continuous crossover from the $M_1$ to the T phase without latent heat [12]. This is in contrast to the first order phase transition between the $M_2$ and T phases.

We now turn to the inter-band transitions in the optical conductivity that are a measure of the electronic structure. From Fig. 4, one can immediately see that the optical conductivity, and thus the electronic structure, of the $M_2$ and T phases is nearly the same. This finding is remarkable given that there are obvious differences in the structural and magnetic properties between the two phases, as discussed above. Interestingly, numerous measurements on single crystals and thin films of the $M_1$ phase give almost the same magnitude of the energy gap as we measure in the $M_2$ and T phases [10,13–17]. The optical energy gap is the spectral region with vanishing conductivity. Above the gap, the optical interband transition labeled Δ, is quite rigid across this wide range of $VO_2$ samples.

For a direct comparison to the $M_1$ phase, in Figure 4 we show optical conductivity extracted from the reflectance spectrum of Verleur *et al.* on single crystals [35]. The complex conductivity is not uniquely determined by the reflectance intensity spectrum without knowledge of the reflectance phase. In addition to the optical conductivity reported in Ref. [35], we present an alternative determination of the optical conductivity using the T phase complex conductivity measured here to approximate the value of the $M_1$ reflectance phase shift in the high energy region of the spectrum. Using this constraint leads to an $M_1$ conductivity spectrum with a lower uncertainty than that reported in the original work, where the reflectance phase shift was not measured. The $M_1$ optical gap is nearly the same as that in the $M_2$ and T phases, and similar optical interband features are present in all three phases.

A schematic of the effective electronic structure of the vanadium *d*-bands for the three insulating $VO_2$ phases is shown in the inset of Fig. 4 (b). There are two features of particular note, labeled Δ and $\Delta_\parallel$. The interband transition Δ across the energy gap is centered about 1.2 eV for all three phases and has little polarization dependence. Similarly, $\Delta_\parallel$ occurs around 2.5 eV in all phases for light polarized along the vanadium chains, and is thus ascribed to transitions between the bonding and anti-bonding $a_{1g}$ bands. These can be thought of as the lower and upper Hubbard bands in the Mott picture. The features labeled Ω occur at 3 eV or higher energies and are primarily optical interband transitions between $O_{2p}$ states and the empty vanadium *d*-states. We emphasize that the robustness of the insulating phase band structure, despite the change in lattice structure, is a remarkable result that is not anticipated by conventional band theory.

To investigate this behavior further, we performed *ab initio* hybrid DFT calculations on the three insulating phases with the Heyd-Scuseria-Ernzerhof (HSE) functional [36,37]. Calculated optical conductivities were determined from the imaginary part of the optical dielectric tensor, using the Vienna *Ab initio* Simulation Package (VASP) [38–41] with HSE (screened) exact-exchange fraction α=0.05 and screening parameter μ=0.2. The optical conductivity calculations are for vertical-only transitions (initial and final states are at the same k-point). The theoretical conductivities were broadened by 0.3 eV, except as indicated, to account for quasiparticle lifetime effects not included in HSE. With suitably chosen α, the HSE functional can, in many instances, provide a good description of electronic properties ranging from band to Mott-Hubbard insulators as shown in previous work [42,43]. The percentage α of exact-change in hybrid DFT can be semi-quantitatively related to the value of the Hubbard U parameter in DFT+U, with larger values of α (and U) yielding larger optical gaps [42–44]. Hybrid DFT and DFT+U both provide a mean-field treatment of on-site 3d correlation on the V atoms. Previous $M_1$ and $M_2$ HSE calculations [45] used α=0.25 calculations, which yielded too large band gaps, compared to experiment [46]. The results of our hybrid DFT calculations are presented in Fig. 4 (c) and (d). In agreement with the experiment, we find that the energy values of the inter-band transitions, particularly $\Delta$ across the optical gap, are quite similar for all three phases. DFT+U calculations (U=5.7 eV and J=0.8 eV, using LDAUTYPE=1 in VASP, not shown) yield qualitatively similar results. This insensitivity to the change in lattice structure in all three insulating phases is incompatible with the Peierls picture. It is interesting to note that the $\Delta_{//}$ feature in the raw HSE result is much sharper than in experiment (see Fig. 4). This is indicative of short lifetimes for carriers excited between the bonding and anti-bonding $a_{1g}$ bands in the real system that is not captured in the static HSE theory. Such lifetime broadening is characteristic of significant electron-electron interactions in these orbitals of Mott-Hubbard character. This is additional evidence that the splitting of the $a_{1g}$ bands, and consequently the energy gap, arises from Coulomb correlations. The most recent iteration of DMFT electronic structure calculations finds energy gaps for the $M_1$ and $M_2$ phases that are consistent with our experimental results [47].

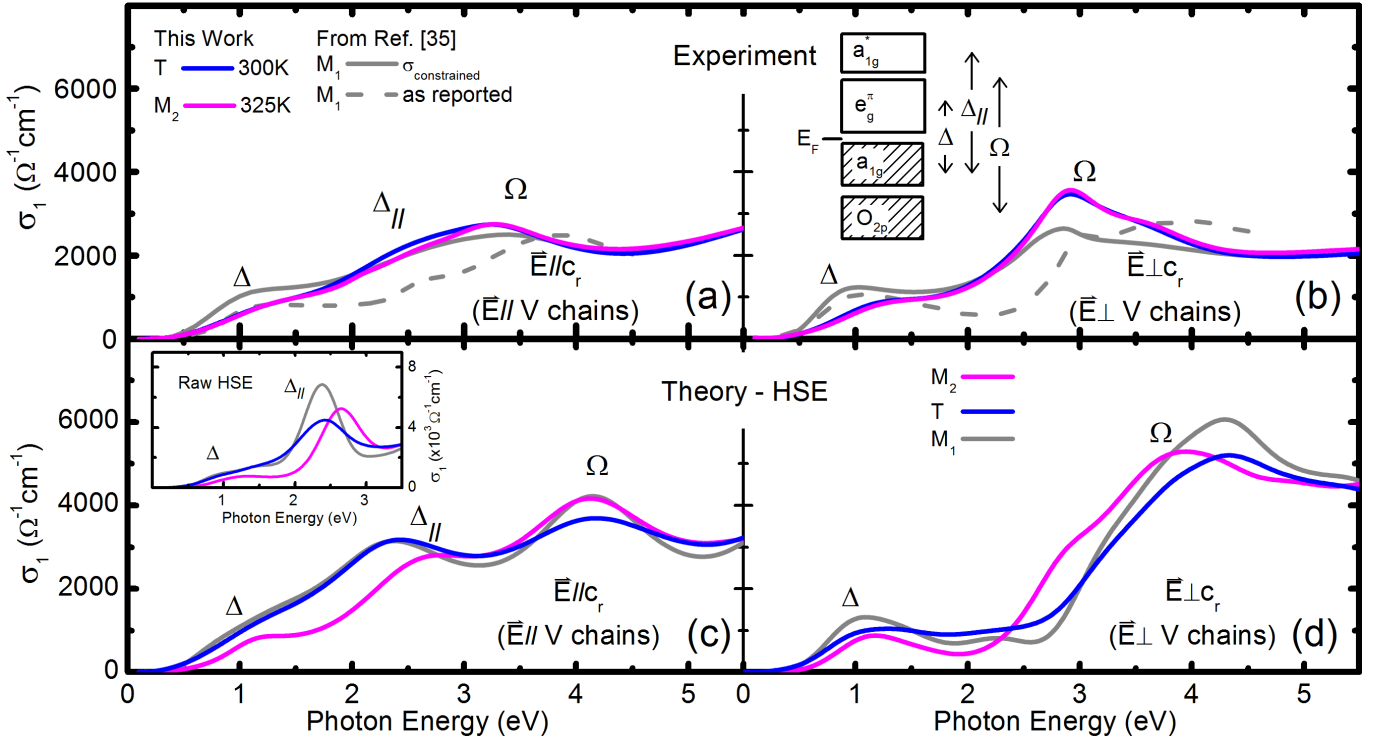

FIG. 4. Experimental optical conductivity $\sigma_1$ of the $M_2$ and triclinic T phases parallel to the vanadium (V) chains (a) and perpendicular to the vanadium (V) chains (b). Phonon features, which occur below 0.11 eV, are not shown. For comparison, accurate optical conductivity of the $M_1$ phase is extracted from the reflectance spectrum reported in [35] by using the complex conductivity of the T phase measured in this work as a constraint above 4 eV (see text). The inset in panel (b) shows an effective energy level diagram along with optical interband transitions that appear in the conductivity spectra. First-principles DFT optical conductivities calculated via the HSE functional are presented in (c) and (d). The calculated "raw" conductivities are broadened by 0.3 eV. The calculated conductivity for $E//c_r$ contains a very sharp $\Delta_{//}$ feature (see inset of panel (c)). To account for lifetime effects not handled in the static HSE treatment, the $\Delta_{//}$ feature, which is assigned to transitions between the lower and upper Hubbard bands in the Mott picture, is further broadened to a FWHM of 1.5 eV in the main panel of (c), which better models the experiment.

To conclude, the nature of the $VO_2$ insulating phases is now clear. The optical spectroscopy data presented in this work clearly demonstrates that the electronic structure of the $VO_2$ insulating phases is robust to changes in lattice structure and vanadium-vanadium pairing. In particular, the energy gap is insensitive to the dimerization of the equally spaced vanadium ions with localized electrons in the $M_2$ chains. This result is incompatible with a Peierls gap and is strong evidence that the gap arises due to Mott-Hubbard type Coulomb correlations. The negative Knight shift is indicative of localized electrons on the equally spaced vanadium ions in the $M_2$ chains. Its absence in the dimerized chains of all three phases [12] elucidates the key subtlety of the insulating $VO_2$ states: in contrast to a more conventional Mott insulator, where valence electrons are localized on individual ions, the dimerized vanadium chains contain bonded spin singlets which are localized on the vanadium dimers. This fact has made it difficult to conclusively distinguish between the Peierls and Mott-Hubbard pictures in the exhaustively studied $M_1$ phase. Study of the $M_2$ and T phases, with their non-equal V chains, is essential to decouple the effects of dimerization and electronic correlations. Seen in a broader context, our work paves a path for disentangling the contributions of the electronic and structural degrees of freedom to phase transitions in other correlated electron systems.

MMQ acknowledges financial support from NSF DMR (grant # 1255156) and the Jeffress Memorial Trust (grant # J-1014). HK acknowledges support from ONR (grant N000141211042) and from the computational facilities at the College of William and Mary. HJ acknowledges support from National Research Foundation of Korea (NRF-2015R1D1A1A01059297). The authors thank Dr. Nobumichi Tamura for discussions on the assignment of twins in the $M_2$ and T phases based on the x-ray diffraction data.  acknowledges support from National Research Foundation of Korea (NRF-2015R1D1A1A01059297). The authors thank Dr. Nobumichi Tamura for discussions on the assignment of twins in the $M_2$ and T phases based on the x-ray diffraction data.

* Corresponding author email address: mumtaz@wm.edu